\documentstyle[epsfig]{article}        
\begin{document}        
\title{On the properties of level spacings \\        
                for decomposable systems}        
\author{Francesco Mezzadri and Antonio Scotti \\        
                 Dept. of Physics, Theoretical Division \\        
                Viale delle Scienze, Univ. di Parma, 43100 Parma, Italy \\     
                 {\em e-mail address:\/} scotti@parma.infn.it }
\date{October 1, 1995}                
\maketitle           
\begin{abstract}        
In this paper we show that the quantum theory of chaos, based on the
statistical theory of energy spectra, presents inconsistencies
difficult to overcome.  In classical  
mechanics a system described by an hamiltonian $H = H_1 + H_2$
(decomposable) cannot be ergodic, because
there are always two dependent integrals of motion          
besides the constant of energy. In quantum mechanics we prove
the existence of decomposable systems \linebreak
$H^q = H^q_1 + H^q_2$ whose spacing distribution agrees
with the Wigner law and we show that in general the spacing distribution       
of $H^q$ is not the Poisson law, even if it has often the same   
qualitative behaviour. We have found that the spacings of $H^q$ are    
among the solutions of a well defined class of homogeneous linear          
systems.  We have obtained an explicit formula for the bases of the kernels  
of these systems, and a chain of inequalities which the coefficients of        
a generic linear combination of the basis vectors must satisfy so that  
the elements of a particular solution will be all positive, i.e. can be
considered a set of spacings.           
\end{abstract}           
           
\section{Introduction}           
\label{first-sec}          
           
When one talks about an ergodic mechanical system in classical mechanics,
everybody  refers to the same definitions and the same theorems. In quantum    
mechanics, unluckily, a universal agreement on quantum chaos doesn't exist.   
Up to now several criteria have been proposed for discriminating chaotic      
hamiltonians from ordered ones, which either present limits and          
inconsistencies difficult to overcome, or need further evidence to have the 
same consistence of the classical 
theory~\cite{Gutzwiller, Paz, Scotti1, Scotti2, Schnirelman, 
Srednicki, Neumann, Zelditch}.  
There is still a lot of confusion  on the subject, so that one could even
doubt  the very  existence of this  vague concept.          
          
The mathematical properties characterising ergodicity in          
quantum mechanics, if they exist at all, are still way ahead of us. Whatever  
they are, however, they ought to be intrinsic, i.e. independent of          
the corresponding classical systems, which, in general, could even not exist.
Notwithstanding this, it is reasonable that when there is a classical limit,   
an ergodicity criterion for quantum systems give answers which agree with the   
classical behaviour~\cite{Scotti2}.  Since in classical mechanics
two non interacting hamiltonians cannot form a chaotic
system, one should ask that this be  true     
also for the quantum counterpart, whatever criterion he is using; otherwise    
serious doubts would rise about its validity.           
           
In the light of this simple test we have analysed
the most popular criterion for quantum chaos, which is
based on the study of the statistical properties 
of energy spectra~\cite{Brody, Gutzwiller, Mehta}:
it is commonly  believed that, if the distribution of
spacings between neighbouring levels obeys  Poisson law,
the system is ordered; otherwise, if the they are distributed  
according to Wigner law, the system is chaotic.           

Clearly the analysis we want to perform is esentially algebraic. Let
the eigenvalues of $H^q$ be           
\begin{equation}           
    E_{i j} = E^1_i + E^2_j \hspace{.5in} i, j = 1,2,  \ldots ,n.       
\label{fund-system}           
\end{equation}           
           
First we have studied the conditions which $E_ {i j}$ must satisfy so that  
the algebraic system~(\ref{fund-system}) can be solved
in the unknowns $E^1_i$ and $E^2_j$; then, carrying out lengthy algebraic     
calculations, we have found that the spacings of $H^q$ are among           
the solutions of a well defined class of homogeneous linear systems.           
We have obtained an explicit formula for the  bases of the kernels of these     
systems, and a chain of inequalities which the coefficients of a generic    
linear combination of the basis vectors must satisfy so that the elements 
of a particular solution will be all positive, i.e. can
be considered a set of spacings.           
           
Solving  these inequalities seems to be a formidable task, and we have not 
yet been able to find a way to do it.   The solution of this problem   
is the starting point for classifying the weights of the possible           
histograms of spacings of decomposable systems. Numerically
we have found histograms  quite
different from Poisson law and also some consistent with the           
Wigner distribution. We think, even if we cannot yet prove it, that their   
weight  is not negligible.           
           
\section{Conditions on the energy levels}           
\label{second-sec}           
           
Eq.~(\ref{fund-system}) in the previous section is a linear system           
of $n^2$ equations in the $2n$ unknowns $E^1_i$ and          
$E^2_j$, whose coefficient matrix has the following structure:           
\begin{equation}           
M = \left( \begin{array}{c |c}          
                S^{(1)}     &     I       \\   \hline           
                S^{(2)}     &     I       \\   \hline           
                          \vdots   &   \vdots             \\   \hline           
                S^{(n)}    &      I              
                      \end{array} \right) .           
\label{coef-matrix}           
\end{equation}           
$I$ is the identity matrix, while $S^{(m)}$ is a matrix           
whose elements are all $0$ except those in the $m-th$ column whose        
entries are all $1$,           
\[    \begin{array}{c}           
                   \hspace{.4in}    m   \\           
       S^{(m)} =     \left(  \begin{array}{ccccc}           
   0  &  \cdots  &  1  &  \cdots  &  0     \\          
   0  &  \cdots  &  1  &  \cdots  &  0     \\           
 \vdots  &  &  \vdots  &  &    \vdots     \\          
   0  &  \cdots  &  1  &  \cdots  &  0     \\          
                                 \end{array}  \right).           
          \end{array}  \]           
           
It is easily seen that the rank of the matrix~(\ref{coef-matrix})            
is $2n - 1$, so the number of dependent equations is           
$n^2 - 2n + 1 = (n - 1)^2$.           
           
It is important to examine carefully the properties of systems whose energy  
levels belong to the range of the linear operator          
$\cal M : \cal R^{2n} \longmapsto \cal  R^{n^2}$, whose matrix          
realization in the canonical basis is given by~(\ref{coef-matrix}).  
           
Let $M_o$ be the $n^2$ by $2n + 1$ matrix           
obtained adding  to $M$ a column vector ${\bf E}$ containing           
the energy levels of the system $H^q = H^q_1 + H^q_2$.  Reducing           
by rows the matrix~(\ref{coef-matrix}), one  obtains           
\begin{equation}           
   \left(  \begin{array}{c |c}           
       U^{(1)}  &  \begin{array}{cccc}           
       1    &    0    &   \cdots   &     0    \\           
       1    &   -1    &   \cdots   &     0    \\           
   \vdots   &  \vdots &  \ddots   &  \vdots   \\          
       1    &    0    &   \cdots   &    -1           
                                           \end{array}   \\  \hline   
     U^{(2)}  &  U^{(1)}   \\  \hline           
                    \vdots       &      \vdots    \\   \hline         
     U^{(n)}  &  U^{(1)}           
  \end{array} \right);           
\label{reduced matrix}           
\end{equation}           
$U^{(m)}$ is a $n$ by $n$ matrix with only one element           
different from zero in the first row:           
\[  U^{(m)}_{ij} = \delta _{1 \; 1 + j - m}.  \]           
Repeating the same sequence of operations leading to
matrix~(\ref{reduced matrix}) on the matrix $M_o$, and imposing that           
it have the same rank of $M$, it can be shown that the elements           
of the vector ${\bf E}$ must satisfy the homogenous linear system           
whose coefficient matrix is the following:           
\begin{equation}           
C = \left(  \begin{array}{c |c |c |c |c}           
    B   &   -B   &   0   &   \cdots   &   0   \\   \hline          
    B   &    0    &  -B  &   \cdots   &   0   \\   \hline           
    \vdots   &   \vdots    &   \vdots   &   \ddots   &   \vdots   \\   \hline 
    B   &   0   &   0   &   \cdots   &   -B            
                 \end{array}  \right),           
\label{hom-matrix}           
\end{equation}          
where $B$ is the matrix
\[  B =  \pmatrix{ 1  &  -1  &  0  &  \cdots &  0  \cr
                  1  &   0  &  -1 &  \cdots &  0  \cr
              \vdots & \vdots & \vdots  & \ddots & \vdots \cr
                  1  &  0  &  0  & \cdots &  -1 \cr}.  \]

The matrix $C$ has $(n - 1)^2$ rows and $n^2$           
columns; it can be decomposed in a natural way in blocks of matrices of 
$n - 1$ rows and $n$ columns. It is reduced by rows; therefore, its           
kernel has dimension $2n - 1$ and coincides, by construction, with the          
range of $\cal M$.  A basis of the kernel obtains  by considering  $2n - 1$   
arbitrary columns of the matrix~(\ref{coef-matrix}). The analitical          
expression of~(\ref{hom-matrix}) is           
\[  B_{kl}=  \delta _{1 \, l} - \delta_{ k + 1 \, l},  \]           
\[  C_{ \lambda \mu \, kl} =          
    \left( \delta_{1 \, \mu} - \delta_{ \lambda + 1 \, \mu} \right) B_{kl}, \]  
\[  \lambda, k = 1,2, \ldots , n-1 ,  \]           
\[  \mu,  l = 1, 2, \ldots ,n . \]           
The indices $\lambda, \mu$ denote  the position of a block in the
matrix $C$, while  $k, l$ locate the single element.  They           
are related to the usual indices of rows and columns $r, s$ by the relations   
\[  r = (\lambda - 1)(n - 1) + k,  \]           
\[  s = (\mu - 1) n + l.  \]          
          
\section{Connection between energy levels and spacings}           
\label{third-sec}           
          
In the previous section we have treated the decomposability of a spectrum;     
yet the statistical analysis concerns the spacings. Our goal is to know   
when, given the number of elements in each cell of a histogram, we can  
find $n^2 - 1$ spacings \linebreak
--- compatible with the histogram --- which may belong to a
decomposable system.          
          
A set of $n^2 -1$ spacings can be consistent with          
$(n^2 - 1)!$, generally distinct, spectra (just this particular 
feature might bring about doubts on the effectiveness of the spacing           
analysis as a tool for probing quantum systems).  We can create a different     
system of levels, apart from an unimportant costant, for every single    
sequence of spacings; without loss of generality, we will set the arbitrary   
constant, which coincides with the eigenvalue of the fundamental state, 
equal to zero in every physical system: in $H^q_1$, $H^q_2$        
and in the total hamiltonian $H^q = H^q_1 + H^q_2$.        
          
With the above convention, a possible system of $n^2$ levels is           
obtained applying the following $n^2$ by $n^2 - 1$ matrix to a vector
of spacings:          
\begin{equation}          
  T = \pmatrix{0  &  0  &  \cdots  &  0  \cr          
                        1  &  0  &  \cdots  &  0  \cr          
                        1  &  1  &  \cdots  &  0  \cr          
                \vdots  &  \vdots  &  \ddots &  \vdots  \cr           
                    1  &  1  &  \cdots  &  1 \cr}.         
\label{connection-matrix}          
\end{equation}          
For  ${\bf E} = T {\bf  \Delta}$, where ${\bf \Delta}$ is the spacing          
vector, the decomposability condition becomes         
\begin{equation}         
 CT{\bf  \Delta} = 0.         
\label{dec-condition}         
\end{equation}         
         
$T$ is not the most general matrix which may connect a set of spacings          
with the levels of a decomposable system; it follows from what
has been stated above that, if one  
arbitrarily permutes the columns of~(\ref{connection-matrix}),  
one obtains a new matrix $T_c$ which connects the same set of spacings with  
another system of levels among the possible $(n^2 - 1)!$.     
Furthermore by the structure of the matrix~(\ref{connection-matrix}) it is    
plain that the elements $E_j$ obtained through $T$ --- or $T_c$ ---         
are increasing with the index $j$.  Given the structure of the 
matrix~(\ref{coef-matrix}), and  our convention,  
the elements of vector ${\bf E}$ are increasing with the index $j$ if        
and only if the spectrum of the second hamiltonian is entirely contained        
within every spacing of the first one, i.e. if they look like a ladder        
``sandwich''.        
        
\begin{figure}
\centerline{\epsfig{file=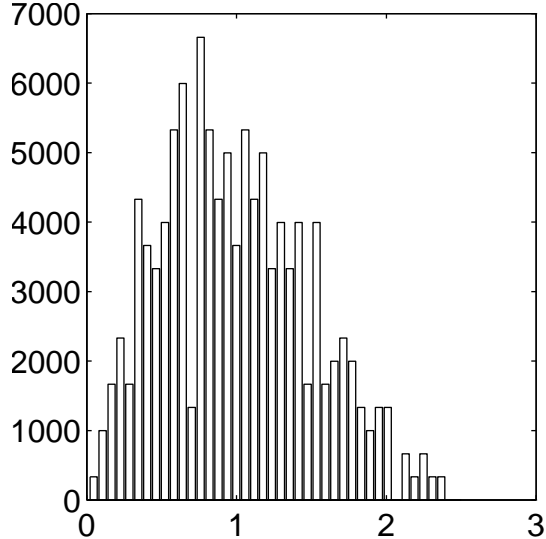}}
\caption{Distribution of spacings of a decomposable sandwich system.}
\label{sand-sys}
\end{figure}

This is, of course, a very particular case, but it is indicative, we believe,
of the possibilities that may occur: as a matter of fact we
can always choose $H^q_1$ and $H^q_2$ so        
that the distribution of spacings of $H^q$ will be
consistent with an arbitrary one:
Poisson, Wigner or whatever we like.  The histogram of spacings of   
$H^q$ is obtained, but for $n$ elements, multiplying by $n - 1$ the number of 
spacings in the cells of the histogram of $H^q_2$, and it will have the        
same qualitative behaviour (see fig.~\ref{sand-sys}).
        
A matrix connecting a decomposable vector ${\bf E}$, whose elements        
are not increasing, is obtained permuting the rows
of~(\ref{connection-matrix}), but, as we shall see, only the pemutations of a
particular class are allowed for rows.        
        
If, without loss of generality, we arrange the levels of $H^q_1$ and        
$H^q_2$ in increasing order, the eigenvalues of the total system must        
obey the following inequalities:        
\begin{equation}        
    E_{i \, j + 1} \ge E_{ij}, \hspace{.5in} E_{i + 1 \, j} \ge E_{ij};        
\label{first-ineq}        
\end{equation}        
the double indices have the same meaning as in eq.~(\ref{fund-system}).     
        
Let us take a decomposable system of levels, not necessarily       
sandwich, and arrange the eigenvalues in increasing order,        
\begin{equation}       
 E_i > E_j \hspace{.3in} \mbox{if} \hspace{.3in}  i > j;        
\label{sec-ineq}        
\end{equation}        
such a vector belongs to the range $K$ of $\cal M$, or to the kernel of   
the matrix~(\ref{hom-matrix}), only if the       
system is sandwich: in all other cases an appropriate permutation       
exists such that the vector $E_{a(i)} \in K$. Inequality~(\ref{sec-ineq})
implies that the elements of the new vector must       
preserve the original order, i.e.      
\begin{equation}      
  E_{a(i)} > E_{a(j)} \hspace{.3in} \mbox{if} \hspace{.3in} i > j.      
\label{third-ineq}     
\end{equation}     
     
A level of the total system can be denoted in two ways: with a double      
index --- $E_{ij}$ --- which shows that it is the sum of the $i-th$ level of  
the first system and the $j-th$ level of the second one, or with the usual   
single index that locates the element in  vector ${\bf E}$. Therefore it
is natural to single out the following association:     
\begin{equation}     
   \begin{array}{ccc}     
              (1,1)  &  \rightarrow  &  1   \\     
              (1,2)  &  \rightarrow  &  2   \\     
            \vdots  &   \vdots   &  \vdots   \\     
              (1,n)  &   \rightarrow  &   n   \\   \hline     
              (2,1)  &  \rightarrow   &  n + 1  \\     
              (2,2)  &   \rightarrow   &  n + 2   \\     
            \vdots  &   \vdots     &   \vdots  \\     
             (2,n)   &  \rightarrow   &  2n   \\   \hline     
            \vdots  &   \vdots     &   \vdots  \\  \hline     
             (n,1)  &   \rightarrow   & n(n - 1) + 1  \\     
             (n,2)   &  \rightarrow   &  n( n - 1) + 2  \\     
            \vdots  &  \vdots      &  \vdots    \\      
             (n,n)   &  \rightarrow  &    n^2     
   \end{array}.     
\label{perm-map}     
\end{equation}

This map will work for sandwich systems only; for a generic      
decomposable system we have to replace the second column with an      
appropriate permutation.  The homogeneous system~(\ref{dec-condition}) 
now becomes     
\begin{equation}         
      CT_p{\bf  \Delta} = 0,         
\label{dec2-condition}         
\end{equation}         
where $T_p$ is obtained from matrix~(\ref{connection-matrix}) by means    
of a suitable permutation of  rows.  In general with an arbitrary     
permutation the system~(\ref{dec2-condition}) may not admit physical     
solutions, i.e. the spacings could be negative, or, equivalently, the system
$C{\bf E} = 0$ doesn't have solutions verifying 
inequalities~(\ref{first-ineq}).  The problem of locating the allowed
permutations is equivalent to finding a particular class of maps \linebreak    
$\cal A^{(p)} : \cal N^2 \longmapsto \cal N$.    
    
It is useful to denote the range of $\cal A^{(p)}$ by means of a $n$ by $n$     
matrix $A^{(p)}$.  From~(\ref{third-ineq}) and~(\ref{perm-map}) it     
follows that a permutation is allowed if and only if    
\[  A_{i \, j + 1}^{(p)} > A_{ij}^{(p)} \hspace{.3in} \mbox{ and }     
\hspace{.3in} A_{i + 1 \, j}^{(p)} > A_{ij}^{(p)};  \]    
this subclass of allowed permutations in  $S_{n^2}$  is made up by all     
and only those  maps which can be allocated 
in a matrix in such a way that the elements     
of the rows are strictly increasing from left to right and the elements of     
the columns are strictly increasing from top to bottom.  It is well known     
that the number of permutations having this property is the     
dimension of the irreducible representation characterized by the 
corresponding square Young tableau.    
    
It follows from the above considerations that the decomposability of a 
nonnegative vector ${\bf \Delta}$ is expressed in the most general way by     
the homogeneous systems~(\ref{dec2-condition}), in which matrix     
$T_p$ is obtained from $T$ by means of a generic permutation of the columns     
and an allowed one of the rows.    
    
It is important to point out, for what  will follow, that the allowed     
permutations leave always the first and the last element fixed, so the     
elements of the first row of $T_p$ are all zero.    
    
A detailed study of the kernels of all matrices $CT_p$ is necessary in order
to classify all possible histograms of decomposable systems.    
   
\section{The chain of inequalities}   
\label{fourth-sec}   
   
The solution of the system~(\ref{dec2-condition}), to make sense, must
be made up of nonnegative elements.    
It follows that in our analysis we have to    
consider only the intersection of the kernel $K_p$ of the  matrix   
$D_p = CT_p$ with the positive hyperoctant of $\cal R^{n^2 -1}$.  An    
arbitrary basis of a subspace is not, in general, formed by vectors whose    
elements are all nonnegative; moreover it is difficult to know a priori if a    
nonnegative vector exists in such a subspace, and, if it does, it is much more  
difficult to find it.  Nevertheless it is easy to prove that if a vector    
of $K_p$ belongs to the positive hyperoctant (origin and coordinate    
hyperplanes excluded) then a whole basis made up of nonnegative vectors    
exists.   
   
Let us find a basis of $K_p$.  Let $\tilde{T}$ be a triangular matrix    
obtained removing the first row from matrix~(\ref{connection-matrix}),
and let $\tilde{T}_p^{-1}$ be the matrix    
obtained permuting the rows and the columns of  $\tilde{T}^{-1}$   
by means of the permutations applied to the columns and to the    
rows of~(\ref{connection-matrix}), respectively, to obtain $T_p$.
The allowed permutations contain $n^2$ elements, while the square matrices    
$\tilde{T}_p^{-1}$ have dimension $n^2 - 1$.  This however is not a    
problem because the first row has been removed from the matrix $T$, and the    
allowed permutations do not change the first element; therefore to obtain    
the right permutation to apply to the columns of    
$\tilde{T}^{-1}$ it is enough to remove the first element of the    
corresponding allowed permutation and to subtract one from the other 
elements. So the matrix $\tilde{T}_p$ coincides with $T_p$ where
the first row has been removed.   
   
We have already noted that the kernel of~(\ref{hom-matrix})
coincides with the range of~(\ref{coef-matrix}).  A basis of 
this subspace is made up of $2n - 1$ arbitrary columns of the matrix $M$. 
Let $\tilde{M}$ be the matrix obtained from $M$ removing the first row: the    
matrix   
\begin{equation}   
T_p \tilde{T}_p^{-1} \tilde{M}   
\label{first-pass}   
\end{equation}   
concides with $M$, except for the first row which now has every element    
equal to zero.  Let $N$ be the submatrix of $\tilde{M}$ obtained    
removing the first and the $(n + 1)-th$ column; the matrix   
\begin{equation}   
W =  \tilde{T}_{p}^{-1} N   
\label{second-pass}   
\end{equation}   
is made up of columns which, by construction, are vectors belonging to   
$K_p$; since $\tilde{T}_p^{-1}$ is obviously invertible,  such   
vectors are linearly independent.  We have excluded the first and the     
$(n + 1)-th$ column of the matrix $\tilde{M}$, because their images   
through $T_p \tilde{T}_p^{-1}$ are the only columns
of~(\ref{first-pass}) which differ from the corresponding columns
of $M$, and therefore do not belong to the kernel 
of~(\ref{hom-matrix}).  It is an easy 
exercise to prove that the vectors of $K_p$
so obtained are a maximal set of linearly  
independent vectors, i.e. that they are a basis.  
  
The kernels of matrices $D_p$ are subspaces changing for different 
choices of the permutations applied to the rows and to the columns 
of~(\ref{connection-matrix}): what we are interested in is a non-empty
intersection of $K_p$ with the positive hyperoctant, it should be clear
by now, that a permutation of the rows of~(\ref{connection-matrix}) can shift
this intersection outside the positive hyperoctant, whereas a permutation
of the columns can only shift it inside.  
  
Now we will find the conditions which must be satisfied by the   
coefficients of an arbitrary linear combination of the columns
of~(\ref{second-pass}) so that such a vector will be nonnegative; 
obviously we will restrict our analysis to matrixes $T_p$ whose rows
are changed by allowed permutations.  
  
The matrix $\tilde{T}^{-1}$ has the following structure:  
\begin{equation}  
   \pmatrix{ 1  &  0  &  \cdots  &  0  &  0  \cr  
                  -1  &  1  &  \cdots  &  0  &  0  \cr  
                   0  &  -1  &  \ddots &  \vdots  &  \vdots  \cr  
                \vdots  &  \vdots  &  \ddots  &  1 &  0  \cr  
                  0  &  0  &  \cdots  &  -1  &  1  \cr };  
\label{inv-matrix}  
\end{equation}  
its analitical expression is the following:  
\[  \tilde{T}_{ij}^{-1} = \delta_{ij} - \delta_{i - 1 \, j};  \]  
the analitical expressions of the columns of the matrix $N$ are the   
following\footnote{Because these are basis vectors, we find
it convenient to use a notation in which the upper index denotes the 
$k-th$ vector, and the lower index the $i-th$ element of such a vector.}:  
\begin{equation}  
  m_i^r = \sum_{k = 1}^n \delta_{i \, n(k - 1) + r} \hspace{.3in}  
      r = 1, 2, \ldots, n-1,  
\label{an-exp1}  
\end{equation}  
\begin{equation}  
m_i^{n-1+r} = \sum_ {k = 0}^{n - 1} \delta_{i \, nr + k}  
\hspace{.3in}  i = 1, 2, \ldots, n^2 - 1. 
\label{an-exp2}  
\end{equation} 
 
Let $\tilde{T}_p^{-1}$ be obtained permuting the columns of  
$\tilde{T}^{-1}$ by means of an allowed permutation; multiplying on the  right  
$\tilde{T}_p^{-1}$  by $N$ we have 
\begin{equation} 
\tilde{T}_p^{-1} N = \tilde{T}^{-1} N_{pi}, 
\label{perm-inv} 
\end{equation} 
where $N_{pi}$ is obtained permuting the rows of $N$  
with the inverse permutation of that applied to the columns of  
$\tilde{T}^{-1}$.  Now formulae~(\ref{an-exp1}) and~(\ref{an-exp2}) become 
\begin{equation}  
  m_i^{r} = \sum_{k = 1}^n \delta_{a^{-1}( i ) \; n(k - 1) + r}  
\hspace{.3in}   r = 1, 2, \ldots, n-1,  
\label{an-exp3}  
\end{equation}  
\begin{equation}  
m_i^{n-1+r} = \sum_ {k = 0}^{n - 1} \delta_{a^{-1}( i ) \; nr + k}  
\hspace{.3in}  i = 1, 2, \ldots, n^2 - 1; 
\label{an-exp4}  
\end{equation} 
$a^{-1}( i )$ denotes the inverse permutation of the allowed one. Finally 
we obtain 
\begin{equation} 
W_{ir} = \delta_{a^{-1}( i ) \; nq + r} -
 \delta_{a^{-1} ( i - 1 ) \; np + r} 
           \hspace{.3in}    i = 1, 2, \ldots, n^2 - 1, 
\label{an-exp5} 
\end{equation} 
\begin{equation} 
W_{i \, r + n - 1} = \delta_{a^{-1}( i ) \; nr + \tilde{p}} - 
   \delta_{a^{-1}( i - 1) \; nr + \tilde{q}}  
\hspace{.3in}  r = 1, 2, \ldots, n - 1; 
\label{an-exp6} 
\end{equation} 
$p, q, \tilde{p}\mbox{ and } \tilde{q}$ are indices running from $0$ to  
$n - 1$.  These expressions show that the second member of
equation~(\ref{an-exp5}) is different from zero if the index $r$ coincides
with the remainder of the division by $n$ of  $a^{-1}(j)$ or $a^{-1}(j - 1)$, 
while the second member of equation~(\ref{an-exp6}) is different from 
zero if the index $r$ is equal to the quotient of the same division.  If we 
require that the individual elements of an arbitrary vector, linear combination
of the columns of $W$, be strictly positive, we obtain the following chain of 
$n^2 -1$ inequalities for the $2n - 2$ coefficients of the linear 
combination:
\begin{equation}
C _{R_{a^{-1}( i )}} + \tilde{C}_{Q_{a^{-1}( i ) }} >
 C _{R_{a^{-1}( i-1 )}} + \tilde{C}_{Q_{a^{-1}( i -1) }};
\label{chain}
\end{equation}
$R_{a^{-1}( i )}$ denotes the remainder of the division by $n$ of  
$a^{-1}( j )$, while $Q_{a^{-1}( i )}$ its quotient.  $C_s$ and
$\tilde{C}_s$ are the coefficients of the columns of $W$,
with the convention that they are zero when $s$ is equal to zero.

We can conclude that if the chain~(\ref{chain}) admits solutions the 
kernel of the matrix $D_p$ intersects the positive hyperoctant and the 
solutions of inequalities~(\ref{chain}) characterise completely the 
intersection.

The problem of classifying histograms compatible with such
an intersection is still unsolved.  
In a histogram the choice of the width of the   
cells depends on the average number of elements per cell; in   
unfavourable cases, when the number of samples is small, few cells or few   
elements per cell yield a meaningless histogram, because a   
different choice of the cells changes completely its behaviour; when, on   
the contrary, the number of samples is large a variation of the width of the   
cells, in a reasonable interval, leaves its appearance unchanged.  In what   
will follow we shall consider two histograms distinct if their cells have a   
different number of elements; clearly two different histograms in our   
convention may represent the same distribution, but, given the freedom in the   
choice of the cells and the algebraic nature of the problem, it is difficult   
to find a different approach.  Nevertheless if one could carry out this   
classification one would succeed in identifying all the distributions   
compatible with decomposable systems.  

In our analysis we will choose the unit of the spacings so that the cells   
have unitary width; moreover a histogram will be denoted by  means of   
a set of $n^2 - 1$ integer $\{n_k\}$  such that  
\begin{equation}  
     \sum_k n_k = n^2 - 1,  
\label{sum-spec}  
\end{equation}  
and such that the number of elements per cell $n_k$ is the number of   
spacings between $k - 1$ and $k$.  The elements of a vector
${\bf \Delta}$,
not necessarily decomposable, denote a partition of $n^2 - 1$;   
they are also the coordinates of a point in the positive hyperoctant of
$\cal R^{n^2 - 1}$ which is located in an hypercube whose edges have unitary   
length and whose vertices have integer and nonnegative coordinates.  If   
we permute the elements of ${\bf \Delta}$, keeping fixed the coordinate   
system, the new vector will be, in general, in another hypercube, similar to   
the previous one and still consistent with the given histogram. It is easy to   
prove that the total number of hypercubes compatible with a given   
histogram is  
\begin{equation}  
\frac{(n^2 - 1)!}{n_1! n_2! \ldots n_k!}.  
\label{num-cub}  
\end{equation}  
The cube in the positive hyperoctant defined by the relation  
\begin{equation}  
0 \leq x_i \leq n^2 - 1 \hspace{.3in} i = 1, 2, \ldots , n^2 - 1  
\label{cube}  
\end{equation}  
contains all the hypercubes compatible with histograms   
physically meaningful.  Given a basis of $K_p$ and the solutions of 
inequalities~(\ref{chain}), one may ask which unitary cubes contained in   
the cube~(\ref{cube}) intersect $K_p$, i.e. which histograms are compatible   
with such a permutation.  Clearly it is enough to consider kernels of   
matrices $D_p$ obtained permuting the rows of the 
matrix~(\ref{connection-matrix}) but not its columns: permuting the   
element of a vector ${\bf \Delta}$ is tantamount as permuting the columns   
of~(\ref{connection-matrix}); therefore if a kernel $K_p$ of   
$D_p = CT_p$, where $T_p$ is obtained permuting only the rows of $T$,   
intersects one of the cubes~(\ref{num-cub}), so does any $K_p$ in   
which the columns of $T_p$ have been permuted with an arbitrary   
permutation.  

\begin{figure} 
\centerline{\epsfig{file=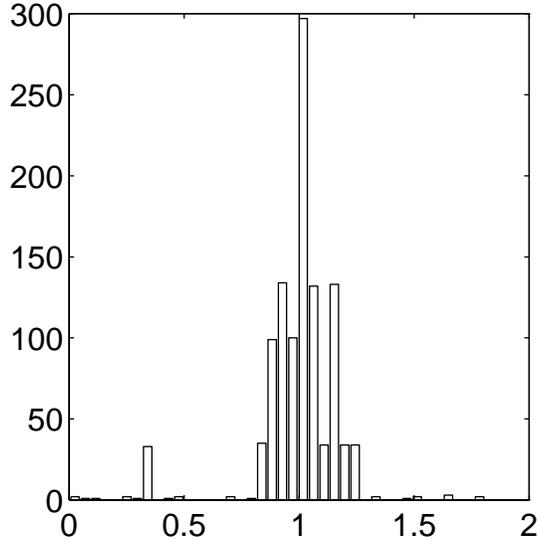}}
\caption{Distribution of spacings of a decomposable non sandwich system.}
\label{tot-sys}
\end{figure}

Solving inequalities~(\ref{chain}) is a formidable task, but even if we   
could do it\linebreak
--- in some simple cases we have been able to solve the   
chain~(\ref{chain}) by means of a computer (see fig.~\ref{tot-sys}) ---,
we would still be faced with the problem of analysing the ensemble of
histograms so generated.

\section{Conclusions}  
\label{fifth-sec}  

The example of the sandwich   
systems and the general study performed above suggest that the weight of   
decomposable systems whose spacing distributions are different from   
Poisson law is not negligible.   The chain of  
inequalities~(\ref{chain}) can be a starting point to give an answer.  
However our results are sufficient evidence that one cannot consider
the distribution of spacings of the levels of a quantum system as the
 "signature " of the so called quantum chaos.    
It is worth to note that the we have not considered the semiclassical limit.   
Even if sometimes the behaviour for $\hbar \rightarrow 0$ can give hints  
on the properties of a system, it can also hide its very quantum nature.  

\section*{Acknowledgements}

We thank Prof. E. Onofri for help and useful discussions.

\end{document}